\documentclass[aps,twocolumn,prc,groupedaddress,longbibliography]{revtex4-1}
\usepackage{amsmath}
\usepackage{graphicx}
\usepackage{color}
\usepackage{here}
\usepackage{bm}

\begin{document}

\title{Large-scale shell-model calculations of nuclear Schiff moments of $^{129}$Xe and $^{199}$Hg}

\author{Kota~Yanase}
\email{yanase@cns.s.u-tokyo.ac.jp}
\author{Noritaka~Shimizu}
\email{shimizu@cns.s.u-tokyo.ac.jp}
\affiliation{Center for Nuclear Study, the University of Tokyo, 7-3-1 Hongo, Bunkyo-ku, Tokyo, Japan}

\date{\today}

\begin{abstract}
The theoretical uncertainty in the nuclear Schiff moment is an obstacle to set constraints on $ CP $ violation beyond the standard model from experimental upper bounds on atomic electric dipole moments.
We perform large-scale shell-model calculations of the $^{129}$Xe and $^{199}$Hg nuclei with realistic effective interactions.
To estimate the Schiff moments caused by the $ P $, $ T $-odd $ \pi NN $ interaction perturbatively, we employ the one-particle one-hole approximation to the intermediate states.
The Schiff moments of $^{129}$Xe and $^{199}$Hg are reduced due to the configuration mixing by $ \sim 10 \% $ from the evaluation of the independent particle model.
On the other hand, the reduction is more significant in mean-field based calculations and shell-model calculations with a drastic truncation.
In order to resolve the discrepancy in the Schiff moment of $^{199}$Hg among several nuclear models, we survey low-energy nuclear structure.
The large-scale shell-model calculations reveal that the Schiff moment of $^{199}$Hg is considerably quenched in the second $ \frac{ 1 }{ 2 }^- $ state.
\end{abstract}

\maketitle

%==========%==========%==========%==========%==========%
%	section: Introduction
%==========%==========%==========%==========%==========%
\section{Introduction \label{intro}}

The permanent electric dipole moments (EDMs) of atoms are expected as probes of charge-parity ($ CP $) violation in beyond the standard model.
It is known that atomic EDMs are greatly enhanced by the relativistic effect of the electron EDM in alkali atoms~\cite{Sandars1965,Sandars1966} and paramagnetic atoms with similar configurations of electron~\cite{Sandars1975}.
It has been demonstrated by atomic many-body calculations that the enhancement factors are greater than 100 in the cesium, thallium, and francium atoms ~\cite{Nataraj2008,Nataraj2011,Mukherjee2009,Ginges2004-review}.
The experimental measurements of the cesium~\cite{Weisskopf1968, Murthy1989} and thallium atoms~\cite{Commins1994,Regan2002} presented upper bounds on the electron EDM.
Recently, an experiment is proposed to measure the atomic EDM of francium in spite of its difficulty owing to the metastability of the nucleus~\cite{Uchiyama2019}.
Recent measurements using the thorium monoxide molecule improved the best limit on the electron EDM by orders of magnitude~\cite{ACME2014,ACME2018}.
The current status is $ | d_{ e } | < 1.1 \times 10^{ -29 } e \, \text{cm} $.

The $ CP $ violation due to the electron EDM must be suppressed in diamagnetic atoms because of the closed configurations of electron~\cite{Flambaum1985-de-eN,Maartensson1987}.
The EDMs of diamagnetic atoms are alternatively sensitive to the $ CP $ violation in atomic nuclei including the $ P $, $ T $-odd nucleon-nucleon ($ NN $) interactions.
The $ P $, $ T $-odd $ NN $ interactions induce nuclear EDMs, but those are completely screened in neutral atoms due to the interactions with surrounding electrons.
One of the leading contributions to the EDM of a diamagnetic atom arises from the nuclear Schiff moment (NSM) induced by the $ P $, $ T $-odd $ NN $ interactions~\cite{Schiff1963,Liu2007-nEDM,Liu2007-NSM}.
The experimental precision has been improved for a long time in $^{199}$Hg~\cite{Lamoreaux1987,Jacobs1993-PRL,Jacobs1995-PRA,Romalis2001,Griffith2009-PRL,Swallows2013-PRA,Graner2016,*Graner2017-erratum} and $^{129}$Xe~\cite{Vold1984,Rosenberry2001,Allmendinger2019,Sachdeva2019}.
In particular the $^{199}$Hg atomic EDM is the most precise measurement among all the particles so far.
The present constraint is $ \left| d_{ \text{Hg} } \right| < 7.4 \times 10^{-30} e \, \text{cm} $.
Some actinide atoms are paid attentions in spite of experimental difficulties because octupole deformation of atomic nuclei is supposed to greatly enhance the NSMs~\cite{Engel2003-225Ra-Schiff,Dobaczewski2018}.
The upper bound of the $^{225}$Ra atomic EDM was first reported several years ago~\cite{Parker2015,Bishof2016}.

In the present study we concentrate on the $^{129}$Xe and $^{199}$Hg NSMs induced by the $ P $, $ T $-odd $ NN $ interactions.
Most of many-body calculations for those nuclei are based on the mean-field approximation so far.
In early studies an independent particle model (IPM) is employed with the phenomenological Woods-Saxon potential and the spin-orbit correction~\cite{Flambaum1985-NN,Flambaum1986}.
Subsequently, the residual interactions were taken into account in the random phase approximation (RPA) and quasi-particle RPA (QRPA)~\cite{Dmitriev2003-PAN,Dmitriev2005_core-polarization,Jesus2005}.
They performed mean-field calculations in $^{198}$Hg and added a neutron to describe the ground state of $^{199}$Hg.
Considering the $ P $, $ T $-odd $ \pi NN $ interaction, the isoscalar and isotensor channels are suppressed, whereas the isovector coefficient is still comparable with results of IPM calculations.
However, fully self-consistent Hartree-Fock-Bogoliubov (HFB) calculations of $^{199}$Hg itself presented controversial results~\cite{Ban2010}.
Although the same Skyrme interactions with the QRPA calculations are employed, the effects of the residual interactions greatly reduce the isovector coefficient so as to change its sign.
The authors inferred that the soft quadrupole deformation of the $^{199}$Hg nucleus might give rise to the theoretical uncertainty and claimed the necessity of configuration mixing.

In the nuclear shell model, wave functions are expressed as linear combinations of the vast number of Slater determinants in the restricted valence space.
In preceding studies, possible configurations are drastically truncated to evade the sizable numerical cost due to a number of active protons and neutron holes of $^{129}$Xe~\cite{Yoshinaga2013,*Yoshinaga2014-NSM-erratum,Teruya2017}.
That simplified version of the shell model is referred to as the pair-truncated shell model (PTSM), in which the many-body bases are composed of collective pairs of like nucleons.
In the PTSM studies, schematic pairing plus quadrupole interactions were adopted.
The RPA ~\cite{Dmitriev2005_core-polarization} and the PTSM~\cite{Teruya2017} studies agree in that the NSM of $^{129}$Xe is reduced roughly by one order of magnitude from the IPM results.

In this paper, we perform large-scale shell-model (LSSM) calculations of $^{129}$Xe and $^{199}$Hg utilizing realistic effective interactions based on the $ G $-matrix interactions.
The $ P $, $ T $-odd $ \pi NN $ interaction can be treated as a perturbation.
As discussed in the following section, we adopt the one-particle one-hole approximation to the intermediate states, where the residual correlations are neglected.
A natural progression of the present study is to expand the valence space in the LSSM calculations.

%==========%==========%==========%==========%==========%
%	section: Formulation
%==========%==========%==========%==========%==========%
\section{Formulation \label{formulation}}

The electric dipole moment (EDM) of an atom is defined by
\begin{align}
 \bm{ d }_{ \text{atom} }
 =
 - \sum_{ i = 1 }^{ Z } e \bm{ r }_{ i }
 ,
\end{align}
where the summation runs over atomic electrons.
The atomic EDM has a non-zero value only if $ P $ and $ T $ symmetries are both violated in the atomic system.
The same argument is applicable to the nuclear Schiff moment (NSM), which requires $ P $ and $ T $ violations in the atomic nucleus.
The NSM violates $ P $ and $ T $ symmetries of the atomic system through the interactions with electrons.

The NSM operator is defined by~\cite{Spevak1997}
\begin{align}
 \bm{ S }
 =
 \frac{ e }{ 10 }
 \sum_{ i = 1 }^{ Z }
 \left(
  r_{ i }^{ 2 } \bm{ r }_{ i }
  -
  \frac{ 5 }{ 3 }
  \left\langle r^{ 2 } \right\rangle_{ \text{ch} }
  \bm{ r }_{ i }
 \right)
 \label{NSM operator}
 ,
\end{align}
where $ \bm{ r }_{ i } $ indicates the proton coordinates with the electric charge $ e $, and $ \langle r^{ 2 } \rangle_{ \text{ch} } $ is the mean squared radius of the charge distribution.
The NSM of a spin-$ J $ state is given by the expectation value in the largest projection $ M = J $.

The NSM can be induced by the $ P $, $ T $-odd $ NN $ interactions.
Considering the one-pion-exchange $ P $, $ T $-odd $ NN $ interaction, the nuclear Hamiltonian contains the following $ P $, $ T $-odd potential:
\begin{align}
 \widetilde{ V }
 & =
 \sum_{ T = 0, 1, 2 }
 \widetilde{ V }_{ T }
 , \notag\\
 \widetilde{ V }_{ 0 }
 & =
 F_{ 0 }
 \big(
  \bm{ \tau }_{ 1 } \! \cdot \! \bm{ \tau }_{ 2 }
 \big)
 \big(
  \bm{ \sigma }_{ 1 } - \bm{ \sigma }_{ 2 }
 \big)
 \! \cdot \!
 \bm{ \nabla }
 \frac{ e^{ - m_{ \pi } r } }{ r }
 , \notag\\
 \widetilde{ V }_{ 1 }
 & =
 F_{ 1 }
 \Big[
  \big(
   \tau_{ 1z } + \tau_{ 2z }
  \big)
  \big(
   \bm{ \sigma }_{ 1 } - \bm{ \sigma }_{ 2 }
  \big)
  \notag\\
  & \qquad \quad
  +
  \big(
   \tau_{ 1z } - \tau_{ 2z }
  \big)
  \big(
   \bm{ \sigma }_{ 1 } + \bm{ \sigma }_{ 2 }
  \big)
 \Big]
 \! \cdot \!
 \bm{ \nabla }
 \frac{ e^{ - m_{ \pi } r } }{ r }
 , \notag\\
 \widetilde{ V }_{ 2 }
 & =
 F_{ 2 }
 \big(
  3 \tau_{ 1z } \tau_{ 2z }
  - \bm{ \tau }_{ 1 } \! \cdot \! \bm{ \tau }_{ 2 }
 \big)
 \big(
  \bm{ \sigma }_{ 1 } - \bm{ \sigma }_{ 2 }
 \big)
 \! \cdot \!
 \bm{ \nabla }
 \frac{ e^{ - m_{ \pi } r } }{ r }
 \label{PT-odd potential}
 ,
\end{align}
where $\bm{ r } = \bm{ r }_{ 1 } - \bm{ r }_{ 2 }$ is the relative coordinate of two nucleons and $ m_{ \pi } $ denotes the pion mass.
The subscripts $ T = 0, 1, 2 $ represent the isospin structures of the $ P $, $ T $-odd vertex $ \overline{ g }_{ \pi N N }^{( T )} $.
The other side of the $ P $, $ T $-odd $ \pi NN $ interaction must be the $ P $, $ T $-even vertex $ g_{ \pi NN } $.
Those couplings are contained in
\begin{align}
 F_{ 0 }
 & =
 \frac{ 1 }{ 8 \pi M_{ N } }
 g_{ \pi N N }
 \overline{ g }_{ \pi N N }^{( 0 )}
 ,
 \notag\\
 F_{ 1 }
 & =
 - \frac{ 1 }{ 16 \pi M_{ N } }
 g_{ \pi N N }
 \overline{ g }_{ \pi N N }^{( 1 )}
 ,
 \notag\\
 F_{ 2 }
 & =
 \frac{ 1 }{ 8 \pi M_{ N } }
 g_{ \pi N N }
 \overline{ g }_{ \pi N N }^{( 2 )}
 \label{F_T def}
 ,
\end{align}
where $ M_{ N } $ denotes the nucleon mass.

In the infinite pion-mass limit, the one-pion-exchange $ P $, $ T $-odd $ NN $ interaction is related to the contact interaction~\cite{Khriplovich-Lamoreaux-CP,Flambaum1986,Dmitriev2005_core-polarization}
\begin{align}
 \widetilde{ V }_{ C }
 =
 \frac{ G }{ \sqrt{ 2 } }
 \frac{ 1 }{ 2m_{ N } }
 \big(
  \eta_{ ab } \bm{ \sigma }_{ a }
  -
  \eta_{ ba } \bm{ \sigma }_{ b }
 \big)
 \! \cdot \!
 \bm{ \nabla }
 \delta ( \bm{ r } )
 \label{eq: contact PT-odd potential}
 ,
\end{align}
where $ G $ is the Fermi coupling constant and
\begin{align}
 \eta_{ nn }
 & =
 \frac{ \sqrt{ 2 } }{ G m_{ \pi }^{ 2 } }
 g_{ \pi N N }
 \big(
  \overline{ g }_{ \pi N N }^{( 0 )}
  +
  \overline{ g }_{ \pi N N }^{( 1 )}
  +
  2
  \overline{ g }_{ \pi N N }^{( 2 )}
 \big)
 \notag\\
 \eta_{ pp }
 & =
 \frac{ \sqrt{ 2 } }{ G m_{ \pi }^{ 2 } }
 g_{ \pi N N }
 \big(
  \overline{ g }_{ \pi N N }^{( 0 )}
  -
  \overline{ g }_{ \pi N N }^{( 1 )}
  +
  2
  \overline{ g }_{ \pi N N }^{( 2 )}
 \big)
 \notag\\
 \eta_{ np }
 & =
 \frac{ \sqrt{ 2 } }{ G m_{ \pi }^{ 2 } }
 g_{ \pi N N }
 \big(
  -
  \overline{ g }_{ \pi N N }^{( 0 )}
  +
  \overline{ g }_{ \pi N N }^{( 1 )}
  -
  2
  \overline{ g }_{ \pi N N }^{( 2 )}
 \big)
 \notag\\
 \eta_{ pn }
 & =
 \frac{ \sqrt{ 2 } }{ G m_{ \pi }^{ 2 } }
 g_{ \pi N N }
 \big(
  -
  \overline{ g }_{ \pi N N }^{( 0 )}
  -
  \overline{ g }_{ \pi N N }^{( 1 )}
  -
  2
  \overline{ g }_{ \pi N N }^{( 2 )}
 \big)
 .
\end{align}
The exchange terms are not contained in the expression~(\ref{eq: contact PT-odd potential}) since nuclear wave functions used in this paper are anti-symmetrized.
The contact $ P $, $ T $-odd $ NN $ interaction originated from the Weinberg operator could be as important as the $ P $, $ T $-odd $ \pi NN $ interaction at low energies~\cite{Dekens2014}.

The nuclear Hamiltonian is expressed as
\begin{align}
 H = H_{ 0 } + \widetilde{ V }
 ,
\end{align}
where $ H_{ 0 }$ denotes $ P $, $ T $-even $ NN $ interactions.
Since the $ P $, $ T $-odd $ \pi NN $ interaction $ \widetilde{ V } $ should be very weak, the ground state of $ H $ is expressed without the normalization as
\begin{align}
 \big| \psi \big\rangle
 =
 \big| \psi_{ 0 } \big\rangle
 +
 \sum_{ n }
 \frac{
  \big| \psi_{ n } \big\rangle
  \big\langle \psi_{ n } \big|
   \widetilde{ V }
  \big| \psi_{ 0 } \big\rangle
 }
 { E_{ 0 } - E_{ n } }
 ,
\end{align}
where $ | \psi_{ 0 } \rangle $ is the ground state of $ H_{ 0 } $ and $ | \psi_{ n } \rangle $ denotes the excited states of the same $ ( J, M = J ) $ and the opposite parity.
The NSM is then calculated as
\begin{align}
 \big\langle \psi \big|
  S_{ z }
 \big| \psi \big\rangle
 & =
 \sum_{ n }
 \frac{
  \big\langle \psi_{ 0 } \big|
   S_{ z }
  \big| \psi_{ n } \big\rangle
  \big\langle \psi_{ n } \big|
   \widetilde{ V }
  \big| \psi_{ 0 } \big\rangle
 }
 { E_{ 0 } - E_{ n } }
 + c.c.
 \notag\\
 & =
 \sum_{ T = 0 }^{ 2 }
 a_{ T }
 g_{ \pi N N }
 \overline{ g }_{ \pi N N }^{( T )}
 \label{Schiff expect}
 ,
\end{align}
where $ c.c. $ denotes the complex conjugate.
The numerical results on nuclear physics are put together in the coefficients $ a_{ T } $, which are referred to as NSM coefficients in this paper.

%----- Figure three types of intermediate states -----%

\begin{figure}[H]
\begin{center}
\includegraphics[width=0.99\linewidth]{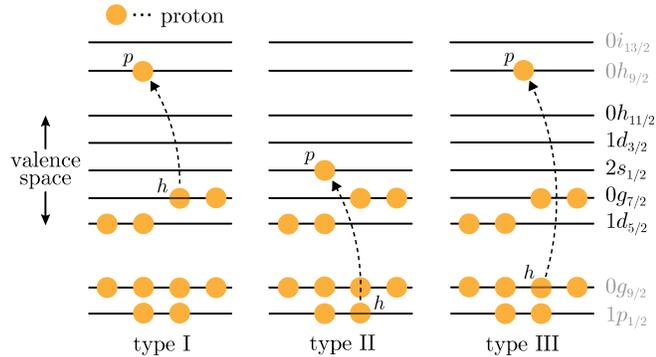}
\caption{
\label{three types of intermediate states}
Three types of the intermediate states defined in Eq.~(\ref{one-particle one-hole excited states}).
As explained in the main text, a proton should be excited across at least one shell gap.
}
\end{center}
\end{figure}

%----------%

In order to obtain wave functions of the ground states in $^{129}$Xe and $^{199}$Hg, we perform large-scale shell-model (LSSM) calculations.
The wave functions of each nucleus are expressed as linear combinations of the possible configurations within a restricted model space referred to as the valence space.
As shown in Fig.~\ref{three types of intermediate states}, the proton valence space consists of five orbitals between magic numbers 50 and 82.
A single-particle orbital in the spherical harmonic oscillator potential with the spin-orbit splitting is characterized by the number of nodes $ n $, the orbital angular momentum $ l $, and the total angular momentum $ j $.
The one-body matrix elements of the NSM operator must vanish unless the initial and final orbitals follow that $ \Delta n \leq  2 $, $ \Delta l = 1 $, and $ \Delta j \leq 1 $.
Since there are no combinations that satisfy all the conditions among the five orbitals in the valence space, we are forced to consider excitations due to the $ P $, $ T $-odd $ \pi NN $ interaction $ \widetilde{ V } $ across at least one shell gap.

In the present calculations, following the prescription of Ref.~\cite{Teruya2017}, the intermediate states $ | \psi_{ n } \rangle $ are approximated by one-particle one-hole excitations from the ground state.
The rotational symmetry of the $ P $, $ T $-odd $ \pi NN $ interaction requires the same spin $ J $ and the projection $ M $ with the ground state.
In the one-particle one-hole approximation, the intermediate states are then expressed with the normalization constants $ N_{ n } $ as
\begin{align}
 \big| \psi_{ n } ; JM \big\rangle
 =
 N_{ n }
 \Big[
  \big[
   c_{ \pi p }^{ \dagger }
   \widetilde{ c }_{ \pi h }
  \big]^{( L )}
  \big| \psi_{ 0 } ; J \big\rangle
 \Big]^{( J )}_{ M }
 \label{one-particle one-hole excited states}
 ,
\end{align}
where the square brackets represent the tensor products.
An intermediate state is specified by single-particle orbitals $ p $ and $ h $ coupled to a rank $ L $.
The one-particle one-hole excitations are classified into three types as illustrated in Fig.~\ref{three types of intermediate states}.
The one-particle one-hole excited states in which protons are excited from the valence space across the $ Z = 82 $ shell gap are referred to as type I.
In type I\hspace{-0.3pt}I excitations, protons are excited from the core to the valence space.
The core orbitals that the excited protons leave should be occupied again in the final state because the core is fully occupied in the shell-model configurations.
The excitations from the core across the valence space are referred to as type I\hspace{-0.3pt}I\hspace{-0.3pt}I.
The details of the calculation of the numerator in Eq.~(\ref{Schiff expect}) are given in Appendix~\ref{app: matele}.

The energy denominators in Eq.~(\ref{Schiff expect}) are approximated by $ E_{ 0 } - E_{ n } \approx \varepsilon_{ h } - \varepsilon_{ p } $, where $ \varepsilon_{ i } $ denotes the single-particle energies in the spherical Nilsson potential.
According to Ref.~\cite{Teruya2017}, the uncertainty arising from this approximation is roughly estimated as $ 10 \% $.

In general, the intermediate states approximated by Eq.~(\ref{one-particle one-hole excited states}) do not compose the orthogonal basis.
The orthogonalization is accomplished by diagonalizing the norm matrix,
\begin{align}
 N_{ n'n }
 =
 \big\langle \psi_{ n' } ; JM \big| \psi_{ n } ; JM \big\rangle
 .
\end{align}
In the present calculations this procedure varies the NSM coefficients $ a_{ T } $ of $^{129}$Xe and $^{199}$Hg within a few percent.

%==========%==========%==========%==========%==========%
%	section: Results
%==========%==========%==========%==========%==========%
\section{Results \label{results}}

%==========%==========%==========%==========%==========%
%	subsection: $^{129}Xe$
%==========%==========%==========%==========%==========%
\subsection{$^{129}$Xe \label{subsec Xe-129}}

For $^{129}$Xe, we adopt the valence space that consists of the five orbitals between the magic numbers 50 and 82, $ 0g_{ 7/2 } $, $ 1d_{ 5/2 } $, $ 1d_{ 3/2 } $, $ 2s_{ 1/2 } $, and $ 0h_{ 11/2 } $, for both proton and neutron.
The $^{129}$Xe nucleus has four protons and seven neutron holes in the valence space and the $ M $-scheme dimension reaches $ 3 \times 10^{ 9 } $.
We utilize the shell-model code KSHELL~\cite{Shimizu2019-KSHELL} and the Oakforest-PACS supercomputer to perform LSSM calculations throughout this paper.

As an effective interaction, we adopt the SN100PN interaction, which is constructed with a renormalized $ G $-matrix derived from the CD-Bonn nucleon-nucleon interaction~\cite{Brown2005}.
The single-particle energies are determined by the low-lying energy levels in $^{133}$Sb and $^{131}$Sn.
The microscopic structure of high-spin states in $^{129}$Xe was investigated by LSSM calculations with the SN100PN interaction~\cite{Kaya2018-131Xe}.
We briefly investigate low-spin structure of $^{129}$Xe, which is relevant to the accuracy of the NSM.

In order to examine the dependence of the $^{129}$Xe NSM on the residual interaction, we employ another well-proven effective interaction.
The SNV interaction consists of the SNBG3 interaction for the neutron-neutron part, the N82GYM interaction for the proton-proton part, and the monopole-based universal ($ V_{ \text{MU} } $) interaction for the neutron-proton part~\cite{Utsuno2014}.
The SNV interaction has been used to investigate nuclear structure of $^{133-134}$Ba and $^{133,135}$La~\cite{Kaya2019-133Ba-134Ba,Laskar2019,Laskar2020}.

%----- Figure 129Xe, spectrum -----%

\begin{figure}[H]
\begin{center}
\includegraphics[width=0.99\linewidth]
{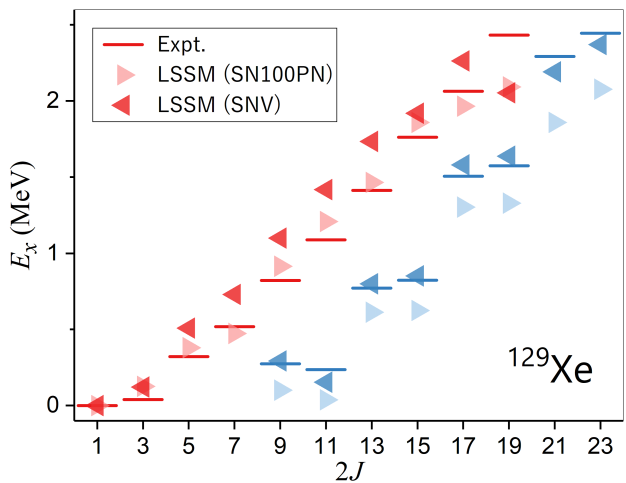}
\caption{
\label{Xe-129 spectrum}
The energy spectra of $^{129}$Xe calculated with the effective interactions SN100PN and SNV.
The positive and negative parity states are represented in red and blue, respectively.
The experimental levels (Expt.) are extracted from Ref.~\cite{NDS129}.
}
\end{center}
\end{figure}

%----------%

Figure~\ref{Xe-129 spectrum} shows the calculated excitation energies of the yrast states in comparison with experimental data~\cite{NDS129}.
The spin and parity $ J^{ \pi } = \frac{ 1 }{ 2 }^+ $ of the ground state are correctly reproduced with both the effective interactions.
Comparing the theoretical results, the positive-parity states with spins $ J \geq \tfrac{ 5 }{ 2 } $ are closer to the experimental levels by using the SN100PN interaction.
Since the negative-parity states are overbound with the SN100PN interaction, the single-particle energy of the neutron $ 0 h_{ 11/2 } $ orbital might be too low for the neutron-deficient nucleus.

%----- Table 129Xe, magnetic moments -----%

\begin{table}[H]
\begin{center}
\caption{
\label{Xe-129 magnetic moments}
Calculated magnetic moments of $^{129}$Xe are compared with experimental data (Expt.) and theoretical results in the PTSM.
The values are given in units of $ \mu_{ N } $.
}
\begin{ruledtabular}
\begin{tabular}{lccc}
	& $ \frac{ 1 }{ 2 }^+_1 $ (g.s.)
	& $ \frac{ 3 }{ 2 }^+_1 $
	& $ \frac{ 11 }{ 2 }^-_1 $ \\[2pt]\hline
LSSM (SN100PN)
	& $ - 0.832 $ & $ 0.590 $ & $ - 1.006 $ \\
LSSM (SNV)
	& $ - 0.858 $ & $ 0.586 $ & $ -1.012 $ \\
Expt.~\cite{NDS129}
	& $ - 0.778 $ & $ 0.58(8) $ & $ - 0.891 $ \\
PTSM~\cite{Higashiyama2011,*Higashiyama2014-PTSM-erratum}
	& $ -0.268 $ & $ 0.278 $ & $ - 1.13 $
\end{tabular}
\end{ruledtabular}
\end{center}
\end{table}

%----------%

Table~\ref{Xe-129 magnetic moments} exhibits the magnetic moments of the ground state and the lowest $ \frac{ 3 }{ 2 }^+ $ and $ \frac{ 11 }{ 2 }^- $ states of $^{129}$Xe.
The spin $ g $-factors are attenuated by a factor of 0.7 from the bare values.
The same effective $ g $-factors $ g_{ s \nu } = -2.68 $ and $ g_{ s \pi } = 3.91 $ were adopted in a PTSM study~\cite{Higashiyama2011,*Higashiyama2014-PTSM-erratum}.
The agreement with the experimental values is considerably improved in the present LSSM calculations.
In particular the quality of the ground-state wave function would be critical to the accuracy of the NSM coefficients.

%----- Table 129Xe -----%

\begin{table}[H]
\caption{
\label{Xe-129 table}
The NSM coefficients of $^{129}$Xe in units of $ 10^{ -2 } e\, \text{fm}^{ 3 } $.
Our final results are given in bold.
}
\begin{ruledtabular}
\begin{tabular}{lccc}
	& $ a_{ 0 } $
	& $ a_{ 1 } $
	& $ a_{ 2 } $
	\\\hline
IPM ($ m_{ \pi } \rightarrow \infty $)
	& $ - 9.9 $
	& $ - 9.9 $
	& $ - 19.8 $
	\\
IPM
	& $ - 4.6 $
	& $ - 4.6 $
	& $ - 9.2 $
	\\
LSSM (SN100PN, $ m_{ \pi } \rightarrow \infty $)
	& $ - 8.7 $
	& $ - 8.2 $
	& $ - 15.8 $
	\\
LSSM (SNV, $ m_{ \pi } \rightarrow \infty $)
	& $ - 8.6 $
	& $ - 8.3 $
	& $ - 16.2 $
	\\
LSSM (SN100PN)
	& $ - 3.7 $
	& $ - 4.1 $
	& $ - 8.0 $
	\\
LSSM (SNV)
	& $ \bm{ - 3.8 } $
	& $ \bm{ - 4.1 } $
	& $ \bm{ - 8.1 } $
	\\[5pt]
IPM ($ m_{ \pi } \rightarrow \infty $)~\cite{Flambaum1985-NN,Flambaum1986}
	& $ - 11 $
	& $ - 11 $
	& $ - 22 $
	\\
IPM~\cite{Dmitriev2005_core-polarization}
	& $ - 6 $
	& $ - 6 $
	& $ - 12 $
	\\
RPA~\cite{Dmitriev2005_core-polarization}
	& $ - 0.8 $
	& $ - 0.6 $
	& $ - 0.9 $
	\\
PTSM~\cite{Yoshinaga2013}
	& $ 0.05 $
	& $ - 0.04 $
	& $ 0.19 $
	\\
PTSM~\cite{Teruya2017}
	& $ 0.3 $
	& $ - 0.1 $
	& $ 0.4 $
\end{tabular}
\end{ruledtabular}
\end{table}

%----------%

Table~\ref{Xe-129 table} summarizes the NSM coefficients $ a_{ T } $ of $^{129}$Xe.
In early studies, Flambaum \textit{et al.} applied the independent particle model (IPM) to calculating the NSM coefficients~\cite{Flambaum1985-NN,Flambaum1986}.
They employed the Woods-Saxon potential with the spin-orbit correction as a mean field.
Dmitriev \textit{et al.} improved the one-body potential by adding a self-consistent mean field obtained from the two-body Landau-Migdal interaction and the Coulomb potential~\cite{Dmitriev2005_core-polarization}.
If the exchange terms in the $ P $, $ T $-odd $ NN $ interactions are excluded, the IPM results follow that $ a_{ 2 } = 2 a_{ 0 } = 2 a_{ 1 } $~\cite{Dmitriev2005_core-polarization} as can be seen in the seventh and eighth rows of Table~\ref{Xe-129 table}.
The IPM is helpful to verify the validity of numerical calculations thanks to the simple relation.
Moreover, IPM results are expected to be insensitive to the one-body potentials as discussed in the following paragraph.

Flambaum \textit{et al.} adopted the contact $ P $, $ T $-odd $ NN $ interaction~\cite{Flambaum1985-NN,Flambaum1986}, whereas the finite-range $ P $, $ T $-odd $ \pi NN $ interaction has been employed in the recent studies.
When the contact interaction was used in Ref.~\cite{Dmitriev2005_core-polarization} for comparison, the IPM results were increased by a factor of 2 from $ a_{ 2 } = 2 a_{ 0 } = 2 a_{ 1 } = - 12 $ obtained by using the finite-range interaction.
Thus, in the limit of the contact interaction, $ m_{ \pi } \rightarrow \infty $, the discrepancy between those IPM results is $ \sim 10 \% $.
This minor difference may come from the improvement of the one-body potential.

The first two rows of Table~\ref{Xe-129 table} show our IPM results.
We construct an artificial configuration in which the proton $ 0g_{ 7/2 }$ and the neutron $ 0h_{ 11/2 } $ orbitals are partially occupied and the last neutron occupies the $ 2s_{ 1/2 } $ orbital.
It is confirmed in the present calculations that the IPM results are increased approximately by a factor of 2 in the infinite pion-mass limit.
Comparing the first and second rows of Table~\ref{Xe-129 table} with the seventh and eighth rows, respectively, it is found that our IPM results are smaller than the IPM results of the early studies by $ 10 - 20 \% $.
The discrepancy could be attributed to the different one-body potentials and occupation probabilities of single-particle orbitals.

The third to sixth rows of Table~\ref{Xe-129 table} show the NSM coefficients by using the wave functions obtained from the LSSM calculations.
The configuration mixing reduces the NSM coefficients by factors of 0.8$ - $0.9 from the IPM results.
It is noticeable that the results are almost independent of the effective interactions, whereas the overlap probability between the ground-state wave functions obtained by using the SN100PN and SNV interactions is 0.85.
On the other hand, those quenching factors are moderate compared with the results in the random phase approximation (RPA)~\cite{Dmitriev2005_core-polarization} and the pair-truncated shell model (PTSM)~\cite{Yoshinaga2013,Teruya2017}, which are shown in the last three rows.

%The results in the infinite pion-mass limit are given in the third and fourth rows of Table~\ref{Xe-129 table}.
%The contact $ P $, $ T $-odd $ NN $ interaction originated from the Weinberg operator could be as important as the $ P $, $ T $-odd $ \pi NN $ interaction at low energies~\cite{Dekens2014}.

In the PTSM, nuclear wave functions are expressed as linear combinations of configurations that are made of collective pairs of like nucleons.
Such a truncation scheme enables the diagonalization of $ H_{ 0 } $ even in an extended valence space~\cite{Yoshinaga2013}.
In the earlier PTSM study, the valence space is enlarged to contain four proton orbitals, $ 2p_{ 1/2 } $, $ 2p_{ 3/2 } $, $ 1f_{ 5/2 } $, and $ 1f_{ 7/2 } $, above the present valence space.
The obtained excited states with a spin-parity of $ \frac{ 1 }{ 2 }^{ - } $ are applied to the intermediate states in Eq.~(\ref{Schiff expect}).
The NSM coefficients are two orders of magnitude smaller than the present LSSM results.
This situation could be reasonable because that approximation of the intermediate states exclude the excitations from the core such as type I\hspace{-0.3pt}I and type I\hspace{-0.3pt}I\hspace{-0.3pt}I excitations in Fig.~\ref{three types of intermediate states}.
The type I contributions should be suppressed since the proton valence space is almost vacant in $^{129}$Xe.
In fact, the type I contributions account for less than one-tenth of the NSM coefficients $ a_{ T } $ in the present calculations.

In the later PTSM study~\cite{Teruya2017}, the valence space is limited to one major shell as the present study, and all the one-particle one-hole excited states are employed as intermediate states.
Although we follow the same prescription, their results are one order of magnitude smaller than the LSSM results.
This inconsistency might be attributed to the drastic truncation scheme and the magnetic moment could be important in reducing the uncertainty of the NSM.
However, the consistency check of the IPM calculations is indispensable for the comparison between the LSSM and PTSM results.
As explained above, our IPM results are effectively consistent with the IPM results in Ref.~\cite{Flambaum1985-NN,Flambaum1986,Dmitriev2005_core-polarization}.
We also confirm that the IPM results for $^{199}$Hg differ from those in preceding studies by at most $ \sim 10 \% $.
We utilize the same computational code to calculate the NSM coefficients by using the wave functions obtained from the LSSM calculations.

%==========%==========%==========%==========%==========%
%	subsection: $^{199}Hg$
%==========%==========%==========%==========%==========%
\subsection{$^{199}$Hg \label{subsec Hg-199}}

We also perform LSSM calculations for $^{199}$Hg and neighboring nuclei.
The neutron valence space consists of the six orbitals between the magic numbers 82 and 126, $ 0h_{ 9/2 } $, $ 1f_{ 7/2 } $, $ 1f_{ 5/2 } $, $ 2p_{ 3/2 } $, $ 2p_{ 1/2 } $, and $ 0i_{ 13/2 } $.
The proton valence space is identical with that of $^{129}$Xe.
As an effective interaction, we adopt the Kuo-Herling interaction, which was originally developed by the $ G $-matrix approach with the use of the Hamada-Johnston potential~\cite{Kuo1972-report,*Herling1972,Mcgrory1975}.
Some of the two-body matrix elements were fitted to the experimental data of $^{206}$Pb, $^{206}$Tl, and $^{206}$Hg~\cite{Blomqvist1984,Rydstrom1990}.

%----- Figure Hg 199, energy spectrum -----%

\begin{figure}[htb]
\begin{center}
\includegraphics[width=0.99\linewidth]
{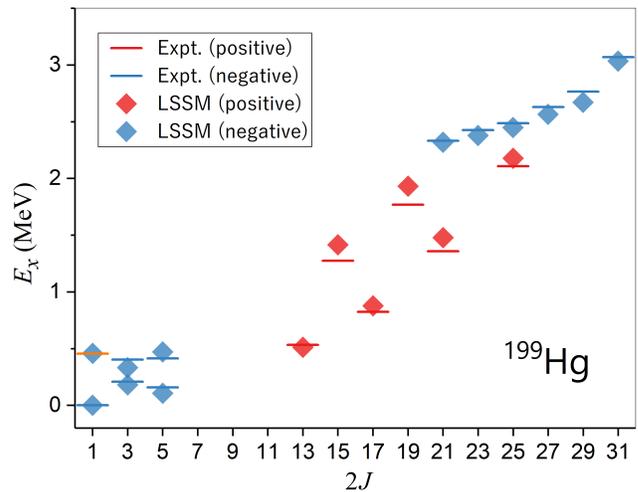}
\caption{
\label{Hg-199 spectrum}
The energy spectrum of $^{199}$Hg.
The experimental data is extracted from Ref.~\cite{NDS199}.
The spin and parity of the experimental level represented in orange are ambiguously assigned $ \frac{ 1 }{ 2 }^- $ or $ \frac{ 3 }{ 2 }^- $.
}
\end{center}
\end{figure}

%----------%

Figure~\ref{Hg-199 spectrum} shows calculated low-lying energy spectrum compared with experimental data.
The correct spin and parity of the ground state are reproduced.
The second $ \frac{ 1 }{ 2 }^- $ state is predicted at 0.458~MeV, which may correspond to the negative parity state discovered in experiment with an excitation energy of 0.455~MeV.
In fact a spin-parity of $ J^{ \pi } = \frac{ 1 }{ 2 }^- $ is assigned as a strong candidate although $ J^{ \pi } = \frac{ 3 }{ 2 }^- $ has not been excluded yet~\cite{Jung1960,Bauer1962,Mathews1975,Lone1975}.

Table~\ref{Hg-199 table} exhibits the NSM coefficients of $^{199}$Hg.
The simple configuration in the IPM is uniquely determined by occupying nucleons from the bottom to the Fermi surface.
The last neutron occupies the $ 2p_{ 1/2 } $ orbital, which gives the correct spin and parity $ J^{ \pi } = \frac{ 1 }{ 2 }^- $.
The first two rows show the results in the IPM.
We confirm the fact that the NSM coefficients are unexpectedly reduced in the infinite pion-mass limit~\cite{Dmitriev2003-PAN,Dmitriev2005_core-polarization}.

As shown in the fourth row, the LSSM results are quenched by 4$ - $12 \% from the IPM evaluations.
The present result of $ a_{ 1 } $ is within the uncertainty estimated in the QRPA calculations with several Skyrme interactions~\cite{Jesus2005}.
Although the same Skyrme interactions are employed, the $ a_{ 1 } $ value is drastically reduced in the fully self-consistent HFB calculations~\cite{Ban2010}.
The results of $ a_{ 0 } $ and $ a_{ 2 } $ are rather close to the HFB results, whereas those values are quenched by one order of magnitude in the QRPA calculations.

%----- Table 199Hg -----%

\begin{table}[H]
\caption{
\label{Hg-199 table}
The NSM coefficients of $^{199}$Hg in units of $ 10^{ -2 } e\, \text{fm}^{ 3 } $.
Our final results are given in the fourth row.
}
\begin{ruledtabular}
\begin{tabular}{lcccc}
	& $ a_{ 0 } $
	& $ a_{ 1 } $
	& $ a_{ 2 } $
	\\\hline
IPM ($ m_{ \pi } \rightarrow \infty $)
	& $ 7.3 $
	& $ 7.3 $
	& $ 14.7 $
	\\
IPM
	& $ 8.4 $
	& $ 8.4 $
	& $ 16.7 $
	\\
LSSM ($ m_{ \pi } \rightarrow \infty $)
	& $ 8.8 $
	& $ 9.2 $
	& $ 19.0 $
	\\
LSSM
	& $ \bm{ 8.0 } $
	& $ \bm{ 7.8 } $
	& $ \bm{ 14.7 } $
	\\
LSSM ($ J^{ \pi } = \frac{ 1 }{ 2 }^-_2$)
	& $ - 0.05 $
	& $ 0.4 $
	& $ 1.3 $
	\\[5pt]
IPM ($ m_{ \pi } \rightarrow \infty $)~\cite{Flambaum1986}
	& $ 8.7 $
	& $ 8.7 $
	& $ 17.4 $
	\\
IPM ($ m_{ \pi } \rightarrow \infty $)~\cite{Dmitriev2003-PAN,Dmitriev2005_core-polarization}
	& $ 5.8 $
	& $ 5.8 $
	& $ 11.6 $
	\\
IPM~\cite{Dmitriev2003-PAN,Dmitriev2005_core-polarization}
	& $ 8.6 $
	& $ 8.6 $
	& $ 17.2 $
	\\
RPA~\cite{Dmitriev2003-PAN,Dmitriev2005_core-polarization}
	& $ 0.04 $
	& $ 5.5 $
	& $ 0.9 $
	\\
IPM~\cite{Jesus2005}
	& $ 9.5 $
	& $ 9.5 $
	& $ 19.0 $
	\\
QRPA~\cite{Jesus2005}
	& $ 0.2 \leftrightarrow 1.0 $
	& $ 5.7 \leftrightarrow 9.0 $
	& $ 1.1 \leftrightarrow 2.5 $
	\\
HFB (SLy4)~\cite{Ban2010}
	& $ 1.3 $
	& $ - 0.6 $
	& $ 2.4 $
	\\
HFB (SkM$^{ * }$)~\cite{Ban2010}
	& $ 4.1 $
	& $ - 2.7 $
	& $ 6.9 $
\end{tabular}
\end{ruledtabular}
\end{table}

%----------%

Here, we present the nuclear spin matrix elements of $^{199}$Hg, which are key information on the contribution from the $ P $, $ T $-odd electron-nucleon interaction to the atomic EDM.
The spin matrix elements of neutron and proton are computed as $ \langle \sigma_{ \nu z } \rangle = -0.322 $ and $ \langle \sigma_{ \pi z } \rangle = -0.006 $, respectively.
Those results support the conclusion from a PTSM calculation~\cite{Yanase2018-199Hg-EDM} that the simple estimate is adequate unless nucleons outside the valence space greatly contribute to the spin matrix elements.

We attempt to exploit a modified Kuo-Herling interaction, which was adjusted further by mainly using high-spin excitation energies of nuclei where the numbers of valence nucleon holes are 2$ - $5~\cite{Szpak2011}.
When the revised effective interaction is employed, the lowest negative parity state of each spin $ J \geq \frac{ 21 }{ 2 } $ is calculated lower than that with the adopted interaction by $ \sim 0.1 $~MeV.
Moreover, the reduced electric quadrupole transition probability, $ B ( E2 ) $, from the $ 8^+_1 $ state to the $ 6^+_1 $ state in $^{200}$Hg is calculated as 7~W.u., which is inconsistent with the experimental value 41(14)~W.u.
Since the $ B ( E2 ; 8^+_2 \rightarrow 6^+_1 ) $ value is calculated as 26~W.u., the lowest $ 8^+ $ state in experiment would correspond to the $ 8^+_2 $ state.
In contrast, low-lying states are little affected by this revision.
The NSM coefficients are reduced by less than $ 2 \% $.

%==========%==========%==========%==========%==========%
%	section: Discussion
%==========%==========%==========%==========%==========%
\section{Discussion \label{discussion}}

In this section, we focus on the serious discrepancies between the theoretical predictions on the NSM coefficients $ a_{ T } $ of $^{199}$Hg.
It was argued in the QRPA study~\cite{Jesus2005} that the neutron excitation from the core plays a crucial role in the destructive interference with $ a_{ 0 } $ and $ a_{ 2 } $.
This effect is excluded in the present framework owing to the restricted valence space, which might be responsible for the relatively moderate quenching in the LSSM calculations.
The $ P $, $ T $-even $ NN $ interactions between the present valence space and the core should be treated explicitly in future studies.
In contrast, the isovector NSM coefficients $ a_{ 1 } $ obtained by the QRPA calculations are comparable with IPM results as shown in Table~\ref{Hg-199 table}.
The dependence of $ a_{ T } $ on the isospin structure of the $ P $, $ T $-odd $ \pi NN $ interaction was explained with the effective potential of the $ P $, $ T $-odd $ \pi NN $ interaction~\cite{Jesus2005}.
As can be seen in Eq.~(9) of Ref~\cite{Engel2003-225Ra-Schiff}, the effective potential does not suppress the isovector NSM coefficient $ a_{ 1 } $, but $ a_{ 0 } $ and $ a_{ 2 } $.
However, the HFB calculations predicted that the isovector NSM coefficient $ a_{ 1 } $ is much more drastically reduced as shown in Table~\ref{Hg-199 table}.
It would be difficult to explain the significant reduction of $ a_{ 1 } $ on the basis of the effective potential.

We discuss a possible origin of the strong dependence of the NSM coefficients $ a_{ T } $ on different nuclear models.
In general low-lying excited states of the same spin and parity could be highly mixed, depending on different nuclear models and effective interactions, with the ground state.
We pay attention to the second lowest $ \frac{ 1 }{ 2 }^- $ state, which is shown in Fig.~\ref{Hg-199 spectrum}.
The NSM coefficients in the $ \frac{ 1 }{ 2 }^-_2 $ state are given in the fifth row of Table~\ref{Hg-199 table}.
It is noticeable that those values are more than one order of magnitude smaller than those in the ground state.
Thus, if the $ \frac{ 1 }{ 2 }^-_2 $ state is mixed with the ground state to some extent in other models, the NSM coefficients will be considerably reduced from the desirable values in the ground state, which are comparable with the IPM evaluation.

In order to understand the strong quenching of the NSM coefficients in the $ \frac{ 1 }{ 2 }^-_2 $ state, we analyze the microscopic nuclear structure.
A significant difference between the lowest two $ \frac{ 1 }{ 2 }^- $ states is the purity of the $ \nu p_{ 1/2 } \otimes 0^+_1 $ configuration, where the ground state of $^{200}$Hg is denoted by $ 0^+_1 $.
In the ground state of $^{199}$Hg, the spectroscopic factor defined by Eq.~(\ref{c2s def}) for the neutron $ 2p_{ 1/2 } $ orbital is calculated as $ S = 0.93 $, which accounts for $ 41 \% $ of the sum-rule value in Eq.~(\ref{c2s sum rule fin}).
This value is consistent with experimental results of $ S = 0.70 (35) $~\cite{Moyer1972,Mathews1975} and $ S = 1.10 $~\cite{Vergnes1985}.
Since the $ \nu p_{ 1/2 } \otimes 0^+_1 $ configuration is similar to the simple configuration of the IPM, it is reasonable that the quenching of the NSM coefficients is moderate in the ground state.
In contrast, the spectroscopic factor is calculated as $ S = 0.05 $ in the $ \frac{ 1 }{ 2 }^-_2 $ state.
It is remarkable that the ratio of the spectroscopic factors of the lowest two $ \frac{ 1 }{ 2 }^- $ states is comparable with the ratio of the results of each NSM coefficient.

It has been demonstrated that experimental energy spectra of several nuclei in the vicinity of $^{208}$Pb are systematically reproduced with the Kuo-Herling interaction and its modified version~\cite{Szpak2011,Wrzesinski2015,Silvestre1981,Broda2011,Chen1972,Cieplicka2018,Podolyak2009-letter,Podolyak2009-full,Steer2008}.
Those agreements ensure the validity of the effective interaction and the ordering of the lowest two $ \frac{ 1 }{ 2 }^- $ states.
It is known that electromagnetic properties can be used to confirm such assignments thanks to the sensitivity to details of nuclear structure~\cite{Neyens2005}.
The magnetic moment of the ground state in $^{199}$Hg was precisely measured as $ 0.506 \mu_{ N } $~\cite{NDS199}.
In the LSSM calculation, the magnetic moment is calculated with effective spin $ g $-factors quenched by a factor of 0.8 as $ 0.459 \mu_{ N } $.
The theoretical magnetic moment of the $ \frac{ 1 }{ 2 }^-_2 $ state is $ 0.612 \mu_{ N } $.
Since these theoretical values of the $ \frac{ 1 }{ 2 }^-_1 $ and $ \frac{ 1 }{ 2 }^-_2 $ states are close to each other, we cannot conclude their correspondence to the experimental ground state utilizing the magnetic moment.

%----- Table 199Hg, B(E2) -----%

\begin{table}[H]
\caption{
\label{199Hg B(E2)}
$ B( E2 ) $ values in units of W.u.
The effective charges $ e_{ \pi } = 1.5 e $ and $ e_{ \nu } = 0.8 e $ are determined by fitting $ B ( E2 ) $ values in even-even Hg isotopes.
}
\begin{ruledtabular}
\begin{tabular}{cccc}
$^{199}$Hg
	& Expt.
	& LSSM
	& LSSM $ \big( \rightarrow \frac{ 1 }{ 2 }^-_2 \big) $
	\\[1pt]\hline
$ \frac{ 5 }{ 2 }^-_1 \rightarrow \frac{ 1 }{ 2 }^-_1 $
	& $ 17.6 (3) $
	& $ 8.7 $
	& $ 2.8 $
	\\
$ \frac{ 5 }{ 2 }^-_2 \rightarrow \frac{ 1 }{ 2 }^-_1 $
	& $ 4.8 (10) $
	& $ 14.2 $
	& $ 0.9 $
	\\
sum
	& $ 22.4 (10) $
	& $ 22.9 $
	& $ 3.7 $
	\\[5pt]
$ \frac{ 3 }{ 2 }^-_1 \rightarrow \frac{ 1 }{ 2 }^-_1 $
	& $ 16.1 (11) $
	& $ 7.2 $
	& $ 8.1 $
	\\
$ \frac{ 3 }{ 2 }^-_2 \rightarrow \frac{ 1 }{ 2 }^-_1 $
	& $ 8.1 (20) $
	& $ 10.4 $
	& $ 3.1 $
	\\
sum
	& $ 24.2 (20) $
	& $ 17.6 $
	& $ 11.2 $
	\\[5pt]
$^{200}$Hg
	& Expt.
	& LSSM
	\\[1pt]\hline
$ 2^+_1 \rightarrow 0^+_1 $
	& 24.57 (22)
	& 22.7
	&
\end{tabular}
\end{ruledtabular}
\end{table}

%----------%

%----- Figure Level scheme of Hg-199, Hg-200 -----%

\begin{figure}[htb]
\begin{center}
\includegraphics[width=0.75\linewidth]
{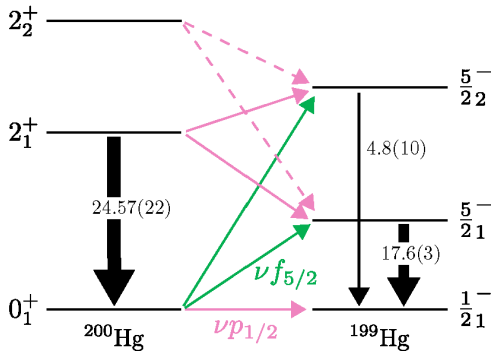}
\caption{
\label{Level-scheme_Hg-199-200}
The predicted main configurations of low-lying states in $^{199}$Hg.
The magenta and green arrows represent the coupling of the neutron $ 2 p_{ 1/2 } $ and $ 1 f_{ 5/2 } $ orbitals to $^{200}$Hg, respectively.
The dashed arrows mean that the $ \nu p_{ 1/2 } \otimes 2^+_2 $ configuration is a minor component in the $ \frac{ 5 }{ 2 }^- $ states.
The experimental $ B ( E2 ) $ values are given in W.u.
}
\end{center}
\end{figure}

%----------%

Table~\ref{199Hg B(E2)} exhibits $ B ( E2 ) $ values between low-lying states in $^{199}$Hg.
The effective charges $ e_{ \pi } = 1.5 e $ and $ e_{ \nu } = 0.8 e $ are determined so that the experimental $ B ( E2 ) $ values of $^{200,202,204,206}$Hg are reproduced in the LSSM calculations.
The $ B ( E2 ) $ enhancement in low-lying excited states of odd-mass nuclei such as $^{199}$Hg was explained with the core excitation model~\cite{deShalit1961}.
In this model, the lowest $ \frac{ 3 }{ 2 }^- $ and $ \frac{ 5 }{ 2 }^- $ states and the second lowest $ \frac{ 3 }{ 2 }^- $ and $ \frac{ 5 }{ 2 }^- $ states are interpreted as admixtures of the $ \nu p_{ 1/2 } \otimes 2^+_1 $ and $ \nu p_{ 1/2 } \otimes 2^+_2 $ configurations~\cite{Kalish1970,Vianden1977}.
However, the present LSSM calculations suggest that the single-particle excitations to the neutron $ 1f_{ 5/2 } $ and $ 2 p_{ 3/2 } $ orbitals are more important than the core excitation to the $ 2^+_2 $ state of $^{200}$Hg.
For example, the lowest two $ \frac{ 5 }{ 2 }^- $ states mainly consist of the $ \nu p_{ 1/2 } \otimes 2^+_1 $ and $ \nu f_{ 5/2 } \otimes 0^+_1 $ configurations as illustrated in Fig.~\ref{Level-scheme_Hg-199-200}.
Thus, the $ \frac{ 5 }{ 2 }^- $ states inherit the collective nature of the $ 2^+_1 $ state of $^{200}$Hg and the $ B ( E2; \frac{ 5 }{ 2 }^- \rightarrow \frac{ 1 }{ 2 }^-_1 ) $ values are enhanced between the $ \nu p_{ 1/2 } \otimes 2^+_1 $ configuration in the $ \frac{ 5 }{ 2 }^- $ states and the $ \nu p_{ 1/2 } \otimes 0^+_1 $ configuration in the ground state.
As shown in Table~\ref{199Hg B(E2)}, the summed $ B ( E2 ) $ values in $^{199}$Hg are very similar to the $ B ( E2; 2^+_1 \rightarrow 0^+_1 ) $ value in $^{200}$Hg.
If the lowest two $ \frac{ 1 }{ 2 }^- $ states were inverted from their experimental order, the summed $ B ( E2 ) $ value of the ground state should be suppressed.

%----- Figure Hg199, NSM accumulation, lowest 1/2^- -----%

\begin{figure}[htb]
\begin{center}
\includegraphics[width=0.99\linewidth]
{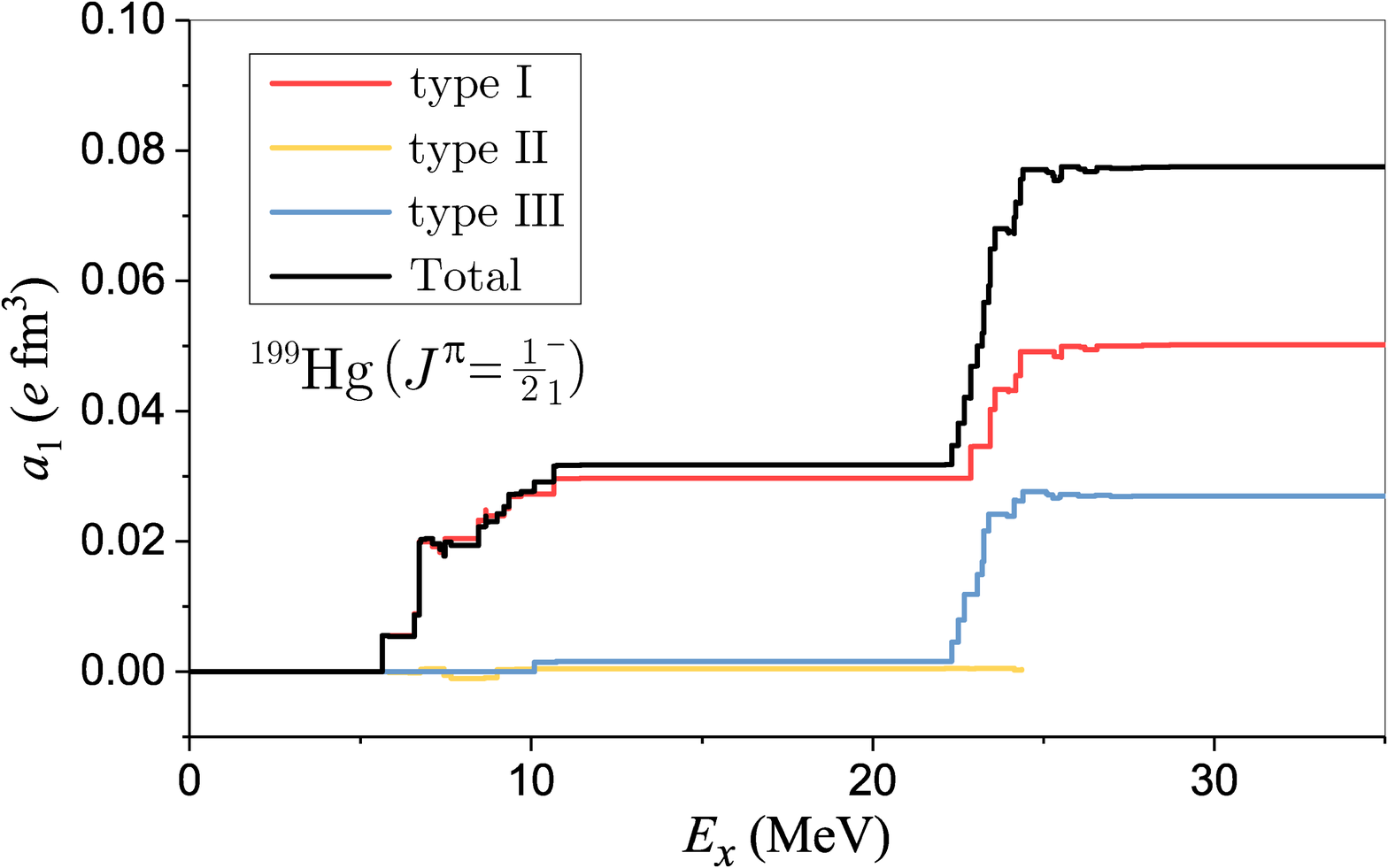}
\caption{
\label{NSM of Hg-199 in 1/2^-_1 accumulation}
The integrated NSM coefficient $ a_{ 1 } $ of $^{199}$Hg in the ground state $ ( J^{ \pi } = \frac{ 1 }{ 2 }^{ - }_{ 1 } ) $.
The horizontal axis represents the excitation energy of the intermediate states.
}
\end{center}
\end{figure}

%----------%

Figure~\ref{NSM of Hg-199 in 1/2^-_1 accumulation} shows the accumulation of the isovector NSM coefficient $ a_{ 1 } $ in the ground state.
The type I\hspace{-0.3pt}I contribution is suppressed in $^{199}$Hg in contrast to $^{129}$Xe because the proton valence space is almost fully occupied.
The type I and type I\hspace{-0.3pt}I\hspace{-0.3pt}I contributions form drastic increases at 7 and 24~MeV.
The NSM operator demands one or three harmonic oscillator excitation, which corresponds to each characteristic excitation energy.
As shown in Fig.~5 in Ref.~\cite{Jesus2005}, such two bumps are also found in the QRPA calculations.
As mentioned above, the residual interactions between the valence space and the core are not included explicitly in the LSSM calculations.
The drawback of the present framework appears in the plateau region between two bumps in Fig.~\ref{NSM of Hg-199 in 1/2^-_1 accumulation}.
Although, the onset of the second bump is more gentle in the QRPA calculations~\cite{Jesus2005}, one can find that the two bumps are separated by an almost flat section except for the result with SI\hspace{-0.3pt}I\hspace{-0.3pt}I Skyrme interaction.

\begin{figure}[htb]
\begin{center}
\includegraphics[width=0.99\linewidth]
{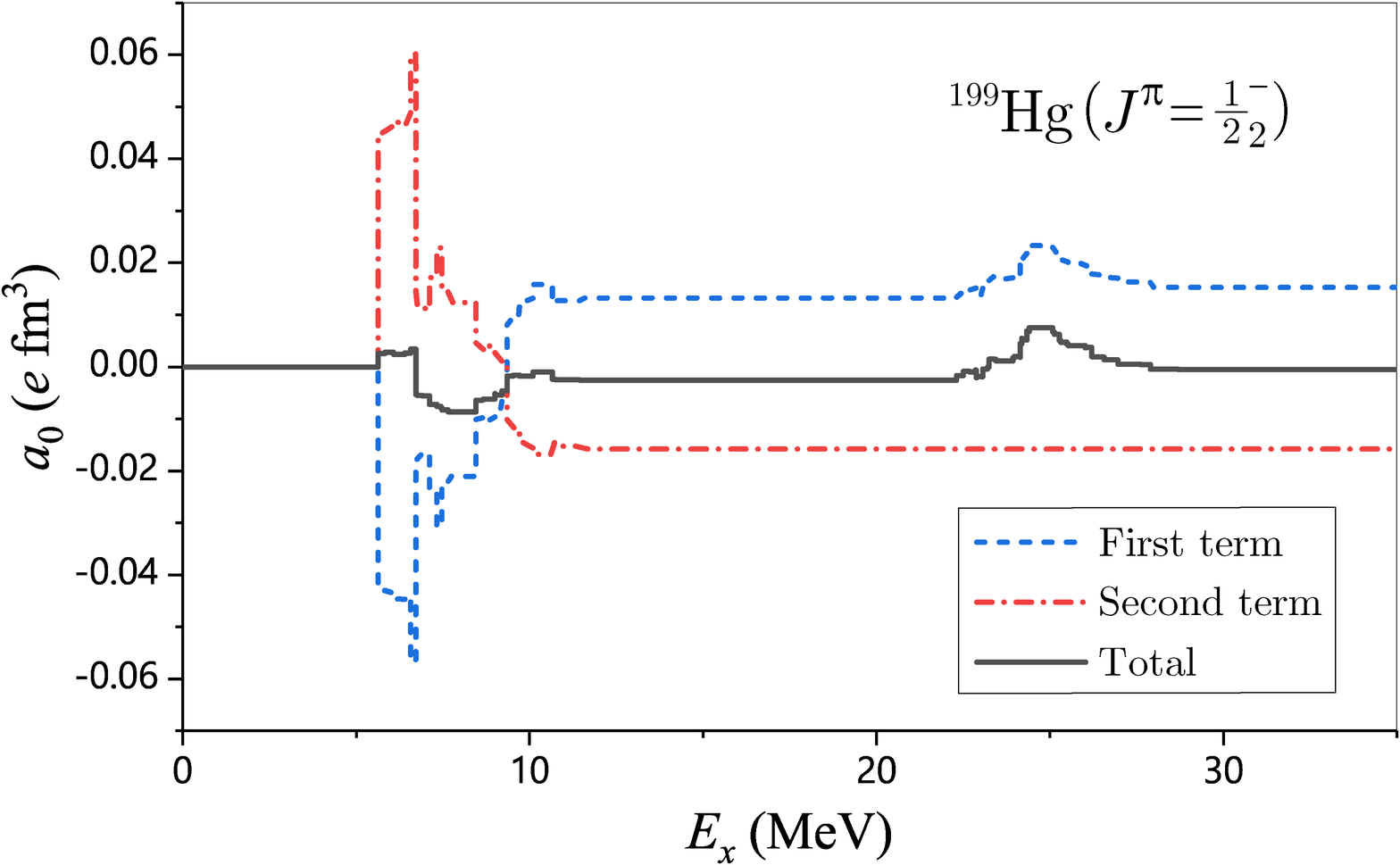}
\caption{
\label{NSM of Hg-199 in 1/2^-_2 accumulation}
The integrated NSM coefficient $ a_{ 0 } $ of $^{199}$Hg in the $ \frac{ 1 }{ 2 }^-_2 $ state.
The contributions from the first term and the second term of the NSM operator are shown in dashed line and dash-dotted line, respectively.
}
\end{center}
\end{figure}

Figure~\ref{NSM of Hg-199 in 1/2^-_2 accumulation} shows the accumulation of isoscalar NSM coefficient $ a_{ 0 } $ in the $ \frac{ 1 }{ 2 }^-_2 $ state.
It is found that the significant reduction results from the cancellation between the contributions from the first and second terms of the NSM operator.
The same feature is found in $ a_{ 0 } $ obtained in the QRPA calculations as shown in Fig.~6 of Ref.~\cite{Jesus2005}.

%----- Figure Hg199, NSM accumulation, second lowest 1/2^- -----%

%----------%

%==========%==========%==========%==========%==========%
%	section: Conclusion
%==========%==========%==========%==========%==========%
\section{Conclusion \label{conclusion}}

We have calculated the nuclear Schiff moments (NSMs) of $^{129}$Xe and $^{199}$Hg using the microscopic wave functions obtained from large-scale shell-model (LSSM) calculations.
The quenching of the NSM coefficients due to the residual interactions is moderate compared with the results in the random phase approximation (RPA), the quasi-particle RPA (QRPA), the fully self-consistent Hartree-Fock-Bogoliubov (HFB) theory, and the pair-truncated shell model except for the isovector NSM coefficient $ a_{ 1 } $ of $^{199}$Hg obtained from the RPA and QRPA calculations.
It has been found that the NSM coefficients in the $ \frac{ 1 }{ 2 }^-_2 $ state of $^{199}$Hg are one order of magnitude smaller than the results in the ground state.
Consequently, if the $ \frac{ 1 }{ 2 }^-_2 $ state is mixed to the ground state by using different nuclear models or effective interactions, the NSM coefficients could be considerably reduced.

The reliability of our calculations is limited by the one-particle one-hole approximation to the intermediate states, whereas the residual correlations might be crucial in the strong quenching of the NSM coefficients.
The residual interactions between the present valence space and the core should be considered in future studies.

Our final result for $^{129}$Xe is given in units of $ e \, \text{fm}^{ 3 } $ as
\begin{align}
 S \left( ^{129} \text{Xe} \right)
 =
 - 0.038 g \overline{ g }^{( 0 )}
 - 0.041 g \overline{ g }^{( 1 )}
 - 0.082 g \overline{ g }^{( 2 )}
 .
\end{align}
The $^{129}$Xe atomic EDM induced by the NSM has been calculated in the Dirac-Fock method~\cite{Dzuba2002,Dzuba2009}, the coupled-perturbed-Hartree-Fock method~\cite{Singh2014,*Singh2014-erratum}, the RPA~\cite{Dzuba2002,Dzuba2009}, and the relativistic coupled-cluster (RCC) model~\cite{Sakurai2019}.
In the recent study using the RCC model, the electric dipole polarizability of the atomic system is accurately reproduced with a discrepancy of $ 2 \% $~\cite{Sakurai2018}.
The state-of-the-art atomic calculation presented $ d_{ A } \, [ e \, \text{cm} ] / S \, [ e \, \text{fm}^{ 3 } ] = 3.20 \times 10^{ -18 } $.
Combining the atomic factor and our result, the $^{129}$Xe atomic EDM is predicted in units of $ e \, \text{cm} $ as
\begin{align}
 d \left( ^{129} \text{Xe} \right)
 & =
 - 1.7 \times 10^{ -18 } \overline{ g }^{( 0 )}
 - 1.8 \times 10^{ -18 } \overline{ g }^{( 1 )}
 \notag\\
 & \quad
 - 3.7 \times 10^{ -18 } \overline{ g }^{( 2 )}
 ,
\end{align}
where $ g_{ \pi NN } = 14.11 $ is adopted~\cite{Yamanaka2016_d-3H-3He-EDM}.
The upper bound in experiment is $ d ( ^{129} \text{Xe} ) < 1.4 \times 10^{ -27 } e \, \text{cm} $~\cite{Sachdeva2019}.

The atomic factor of $^{199}$Hg has been calculated in the Dirac-Fock method~\cite{Dzuba2002,Dzuba2009}, the multi-configuration Dirac-Hartree-Fock method~\cite{Radziute2014,Radziute2016}, the RPA~\cite{Dzuba2002,Dzuba2009}, the configuration interaction method~\cite{Dzuba2009}, and the RCC model~\cite{Latha2009,*Latha2009-note,*Latha2015-erratum,Singh2015,Sahoo2017,Sahoo2018}.
The latest study based on the RCC model presented $ d_{ A } \, [ e \, \text{cm} ] / S \, [ e \, \text{fm}^{ 3 } ] = - 1.77 \times 10^{ -17 } $.
In this model, the electric dipole polarizability is well reproduced.
Combining the atomic factor and our result, which is given in units of $ e \, \text{fm}^{ 3 } $ as
\begin{align}
 S \left( ^{199} \text{Hg} \right)
 =
 0.079 g \overline{ g }^{( 0 )}
 + 0.075 g \overline{ g }^{( 1 )}
 + 0.143 g \overline{ g }^{( 2 )}
 ,
\end{align}
the atomic EDM is calculated in units of $ e \, \text{cm} $ as
\begin{align}
 d \left( ^{199} \text{Hg} \right)
 & =
 - 2.0 \times 10^{ -17 } \overline{ g }^{( 0 )}
 - 1.9 \times 10^{ -17 } \overline{ g }^{( 1 )}
 \notag\\
 & \quad
 - 3.7 \times 10^{ -17 } \overline{ g }^{( 2 )}
 ,
\end{align}
whereas the current limit is given in $ 95 \% $ C.L. as $ d ( ^{199} \text{Hg} ) < 7.4 \times 10^{ -30 } e \, \text{cm} $~\cite{Graner2016,*Graner2017-erratum}.

\begin{acknowledgements}
This research was supported by MEXT and JICFuS as post-K priority issue 9 (hp180179, hp190160) and Program for Promoting Researches on the Supercomputer ``Fugaku'' (Simulation for basic science: from fundamental laws of particles to creation of nuclei).
It was also supported by KAKENHI grant (17K05433).
The numerical calculation was performed mainly on the Oakforest-PACS supercomputer for Multidisciplinary Computational Sciences Project of Tsukuba University (xg18i035).
We acknowledge Cenxi Yuan for the discussions about the Kuo Herling interaction,
 Yutaka Utsuno for the discussions about the spectroscopic factor, Naotaka Yoshinaga, and Koji Higashiyama for helpful discussions.
\end{acknowledgements}

\appendix

%==========%==========%==========%==========%==========%
%	section: Matrix elements
%==========%==========%==========%==========%==========%
\section{Matrix elements \label{app: matele}}

The NSM operator and the $ P $, $ T $-odd $ \pi NN $ interaction are expressed as
\begin{align}
 S_{ z }
 & =
 \sum_{ ij }
  s_{ ij }
  c_{ i }^{ \dagger }
  c_{ j }
 , \quad
 \widetilde{ V }
 =
 \sum_{ i < j }
 \sum_{ k < l }
  \widetilde{ v }_{ ijkl }
  c_{ i }^{ \dagger }
  c_{ j }^{ \dagger }
  c_{ l }
  c_{ k }
 ,
\end{align}
where $ s_{ ij } $ and $ \widetilde{ v }_{ ijkl } $ are the one-body matrix elements of $ S_{ z } $ and the two-body matrix elements of $ \widetilde{ V } $, respectively.
A subscript denotes a single-particle state and the $ z $-component of isospin.
The many-body matrix elements are calculated as follows.

As explained in Sec.~\ref{formulation}, there are three types of the intermediate states.
In type I one-particle one-hole excitations, protons in the valence space are excited to higher orbitals across the $ Z = 82 $ shell gap.
In type I\hspace{-0.3pt}I, protons are excited from the core to the valence space.
In type I\hspace{-0.3pt}I\hspace{-0.3pt}I, protons are excited from the core across the valence space.

%----- Figure variety of diagrams -----%

\begin{figure}[H]
\begin{center}
\includegraphics[width=0.99\linewidth]{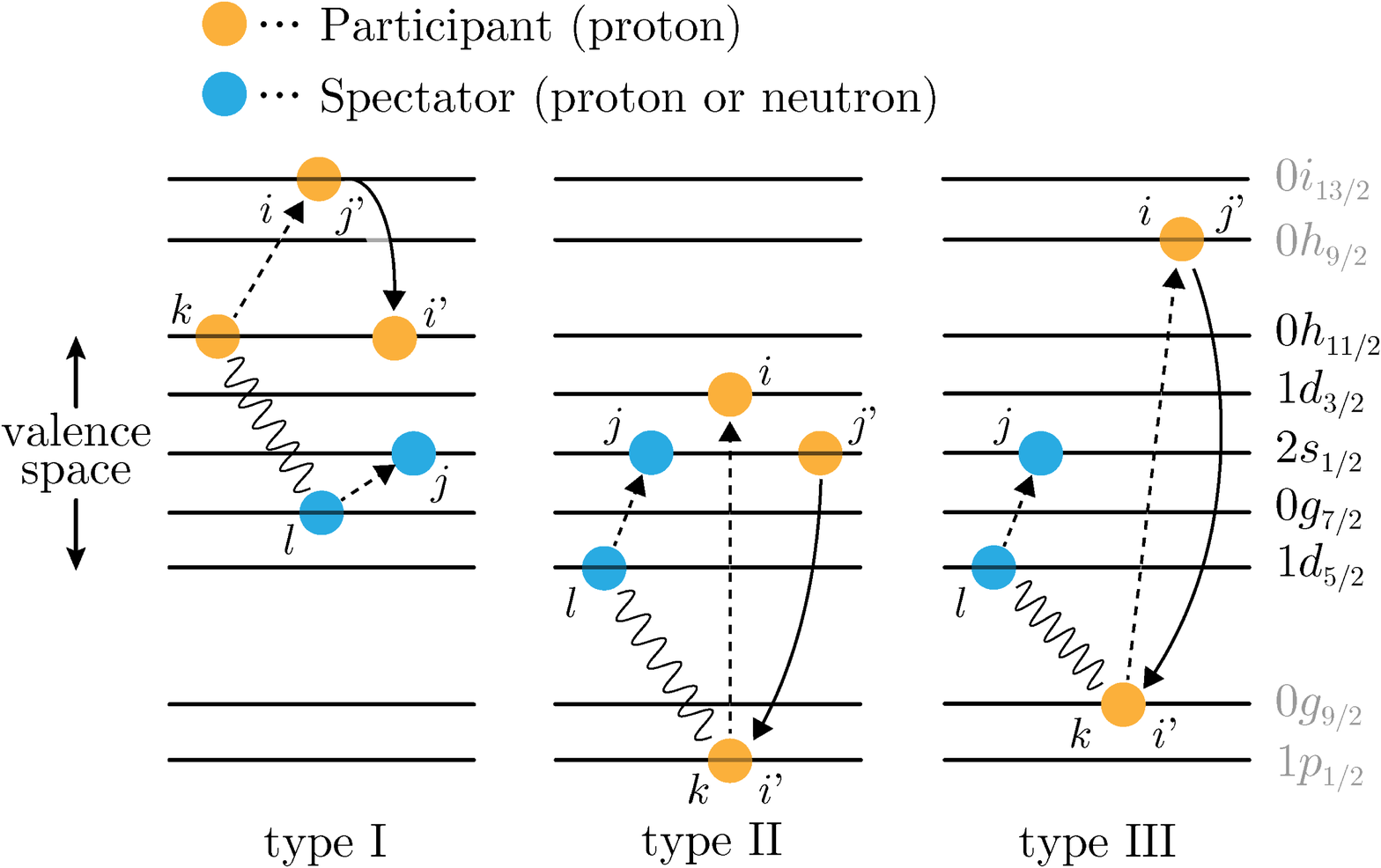}
\caption{
\label{variety of excitation}
A schematic diagram illustrating the one-body and two-body transitions through the NSM operator (solid arrows) and the $ P $, $ T $-odd $ \pi NN $ interaction (dashed arrows), respectively.
The $ P $, $ T $-odd $ \pi NN $ interaction represented by a wavy line should excite a proton, which is referred to as a participant, across at least one shell gap.
The participant proton interacts with a ``spectator'', which is a proton or neutron. Although the spectators are shown in the valence space, nucleons in the core also behave as spectators that remain in the same orbitals ($ j = l $).
The indices indicating single-particle states correspond to those in Eqs.~(\ref{nsm type I})-(\ref{nsm type III}).
}
\end{center}
\end{figure}

%----------%

For type I excitations, the many-body matrix elements in Eq.~(\ref{Schiff expect}) can be reduced to
\begin{align}
 &
 \big\langle \psi_{ 0 } \big|
  S_{ z }
 \big| \psi_{ n } \big\rangle
 \big\langle \psi_{ n } \big|
  \widetilde{ V }
 \big| \psi_{ 0 } \big\rangle
 \notag\\
 &
 =
 \big\langle \psi_{ 0 } \big|
  \Big(
   \sum_{ i'j' }
   s_{ i'j' }
   c_{ i' }^{ \dagger }
   a_{ j' }
  \Big)
  a_{ p }^{ \dagger }
  c_{ h }
 \big| \psi_{ 0 } \big\rangle
 \notag\\
 & \qquad \times
 \big\langle \psi_{ 0 } \big|
  c_{ h }^{ \dagger }
  a_{ p }
  \Big(
   \sum_{ ij } \sum_{ k \leq l }
   \widetilde{ v }_{ ijkl }
   a_{ i }^{ \dagger }
   c_{ j }^{ \dagger }
   c_{ l }
   c_{ k }
  \Big)
 \big| \psi_{ 0 } \big\rangle
 \notag\\
 &
 =
 \sum_{ i'j } \sum_{ k \leq l }
 s_{ i'p }
 \widetilde{ v }_{ pjkl }
 \big\langle \psi_{ 0 } \big|
  c_{ i' }^{ \dagger }
  c_{ h }
 \big| \psi_{ 0 } \big\rangle
 \big\langle \psi_{ 0 } \big|
  c_{ h }^{ \dagger }
  c_{ j }^{ \dagger }
  c_{ l }
  c_{ k }
 \big| \psi_{ 0 } \big\rangle
 \label{nsm type I}
 ,
\end{align}
where $ a_{ i }^{ \dagger } $ is the proton creation operator of a single-particle orbital higher than the valence space.
Here, $ a_{ i } \big| \psi_{ 0 } \big\rangle = 0 $ is used.
For type I\hspace{-0.3pt}I excitations, we have
\begin{align}
 &
 \big\langle \psi_{ 0 } \big|
  S_{ z }
 \big| \psi_{ n } \big\rangle
 \big\langle \psi_{ n } \big|
  \widetilde{ V }
 \big| \psi_{ 0 } \big\rangle
 \notag\\
 &
 =
 \big\langle \psi_{ 0 } \big|
  \Big(
   \sum_{ i'j' }
   s_{ i'j' }
   b_{ i' }
   c_{ j' }
  \Big)
  c_{ p }^{ \dagger }
  b_{ h }^{ \dagger }
 \big| \psi_{ 0 } \big\rangle
 \notag\\
 & \qquad \times
 \big\langle \psi_{ 0 } \big|
  b_{ h }
  c_{ p }
  \Big(
   \sum_{ i \leq j } \sum_{ kl }
   \widetilde{ v }_{ ijkl }
   c_{ i }^{ \dagger }
   c_{ j }^{ \dagger }
   c_{ l }
   b_{ k }^{ \dagger }
  \Big)
 \big| \psi_{ 0 } \big\rangle
 \notag\\
 &
 =
 \sum_{ i \leq j } \sum_{ j'l }
 s_{ hj' }
 \widetilde{ v }_{ ijhl }
 \big\langle \psi_{ 0 } \big|
  c_{ j' }
  c_{ p }^{ \dagger }
 \big| \psi_{ 0 } \big\rangle
 \big\langle \psi_{ 0 } \big|
  c_{ p }
  c_{ i }^{ \dagger }
  c_{ j }^{ \dagger }
  c_{ l }
 \big| \psi_{ 0 } \big\rangle
 \label{nsm type II}
 ,
\end{align}
where the proton hole creation operator of a core orbital $ b_{ k }^{ \dagger } $ follows $ b_{ k } \big| \psi_{ 0 } \big\rangle = 0 $.
For type I\hspace{-0.3pt}I\hspace{-0.3pt}I excitations, we have
\begin{align}
 &
 \big\langle \psi_{ 0 } \big|
  S_{ z }
 \big| \psi_{ n } \big\rangle
 \big\langle \psi_{ n } \big|
  \widetilde{ V }
 \big| \psi_{ 0 } \big\rangle
 \notag\\
 &
 =
 \big\langle \psi_{ 0 } \big|
  \Big(
   \sum_{ i'j' }
   s_{ i'j' }
   b_{ i' }
   a_{ j' }
  \Big)
  a_{ p }^{ \dagger }
  b_{ h }^{ \dagger }
 \big| \psi_{ 0 } \big\rangle
 \notag\\
 & \qquad \times
 \big\langle \psi_{ 0 } \big|
  b_{ h }
  a_{ p }
  \Big(
   \sum_{ ijkl }
   \widetilde{ v }_{ ijkl }
   a_{ i }^{ \dagger }
   c_{ j }^{ \dagger }
   c_{ l }
   b_{ k }^{ \dagger }
  \Big)
 \big| \psi_{ 0 } \big\rangle
 \notag\\
 &
 =
 \sum_{ jl }
 s_{ hp }
 \widetilde{ v }_{ pjhl }
 \big\langle \psi_{ 0 } \big|
  c_{ j }^{ \dagger }
  c_{ l }
 \big| \psi_{ 0 } \big\rangle
 \label{nsm type III}
 .
\end{align}
The one-body and two-body matrix elements in Eqs.~(\ref{nsm type I})-(\ref{nsm type III}) are computed by using the ground-state wave functions $ | \psi_{ 0 } \rangle $ obtained from the LSSM calculations.
Figure~\ref{variety of excitation} shows a schematic explanation of the contribution to the NSM from each type of the intermediate states.

%==========%==========%==========%==========%==========%
%	section: Spectroscopic factor
%==========%==========%==========%==========%==========%
\section{Spectroscopic factor \label{app: c2s}}

The spectroscopic factors for the single-neutron stripping reactions of $^{199}$Hg are defined by~\cite{Ring-Schuck_NMBP}
\begin{align}
 &
 S_{ k } ( n_{ i } , J_{ i } ; n_{ f } , J_{ f } )
 =
 \frac{ 1 }{ 2J_{ f } + 1 }
 \notag\\
 & \qquad \times
 \left|
  \Big\langle {}^{200} \text{Hg} ( n_{ f }, J_{ f } ) \Big| \! \Big|
   c_{ k }^{ \dagger }
  \Big| \! \Big| {}^{199} \text{Hg} ( n_{ i }, J_{ i } ) \Big\rangle
 \right|^{ 2 }
 \label{c2s def}
 ,
\end{align}
where $ k $ is the single-particle orbital of the neutron and $ ( n, J ) $ denotes the $ n $th lowest state with the spin $ J $.
In general $ C^{ 2 } $ is multiplied, where $ C $ is the Clebsch-Gordan coefficient of isospin.
In the present calculations of $^{199}$Hg, it follows $ C = 1 $.
The spectroscopic factors satisfy sum rules as
\begin{align}
 &
 \sum_{ n_{ f } J_{ f } }
 ( 2J_{ f } + 1 )
 S_{ k }
 \notag\\
 & \qquad
 =
 ( 2J_{ i } + 1 )
 \left[
  ( 2j + 1 )
  - N^{ i }_{ k } \left( n_{ i }, J_{ i } \right)
 \right]
 \label{c2s sum rule fin}
 ,
\end{align}
where $ N^{ i }_{ k } $ and $ N^{ f }_{ k } $ are the occupation numbers of a single-particle orbital $ k $ in the initial and final states, respectively.

%==========%==========%==========%==========%==========%
%	section: Sign conventions of the $ P $, $ T $-odd $ \pi NN $ interaction
%==========%==========%==========%==========%==========%
\section{Sign conventions of the $ P $, $ T $-odd $ \pi NN $ interaction \label{app: PT-odd pi NN}}

Table~\ref{Vpt sign} summarizes the sign conventions of $ F_{ T } $ in Eq.~(\ref{F_T def}) and the isospin $ z $-component of neutron, which is involved in the sign of $ \widetilde{ V }_{ 1 } $ in Eq.~(\ref{PT-odd potential}), adopted in preceding works and ours.
We infer the isospin $ z $-component of neutron $ \langle \tau_{ z } \rangle_{ n } = -1 $ adopted by Dmitriev \textit{et al.} from Eq.~(7) in Ref.~\cite{Dmitriev2005_core-polarization}.
This assumption is also supported by the consistency of the IPM results of $ a_{ 1 } $ as  shown in Tables~\ref{Xe-129 table} and \ref{Hg-199 table}.
In this paper we follow the conventions of Refs.~\cite{Jesus2005,Ban2010}.

%----- Table sign of Vpt -----%

\begin{table}[H]
\caption{
\label{Vpt sign}
The sign conventions of $ F_{ T } $ and the isospin $ z $-component of neutron.
$ F_{ T } = -1 $ means the sign convention is opposite to ours.
}
\begin{ruledtabular}
\begin{tabular}{lcccc}
& $ F_{ 0 } $ & $ F_{ 1 } $ & $ F_{ 2 } $ & $ \langle \tau_{ z } \rangle_{ n } $ \\\hline
Engel \textit{et al.}~\cite{Jesus2005,Ban2010} and this paper
	& $ +1 $ 	& $ +1 $ 	& $ +1 $ 	& $ +1 $ 	\\
Domitriev \textit{et al.}~\cite{Dmitriev2003-PAN,Dmitriev2005_core-polarization}
	& $ -1 $ 	& $ +1 $ 	& $ +1 $ 	& $ ( -1 ) $ 	\\
Yoshinaga \textit{et al.}~\cite{Yoshinaga2013,Teruya2017}
	& $ +1 $ 	& $ -1 $ 	& $ +1 $ 	& $ +1 $
\end{tabular}
\end{ruledtabular}
\end{table}

%----------%

%\bibliography{edm,nuclear-structure,nuclear-data-sheets}
\providecommand{\noopsort}[1]{}\providecommand{\singleletter}[1]{#1}%\providecommand{\noopsort}[1]{}\providecommand{\singleletter}[1]{#1}%\providecommand{\noopsort}[1]{}\providecommand{\singleletter}[1]{#1}%

\end{document}